\documentclass[useAMS,usenatbib]{mn2e}
\usepackage{graphicx}
\usepackage{txfonts}

\title[First high energy observations of Narrow Line Seyfert 1s]
  {First high energy observations of Narrow Line Seyfert 1s with INTEGRAL/IBIS}
\author[A. Malizia et al.]
{A.~Malizia,$^1$
  L.~Bassani,$^1$ A.J.~Bird,$^2$ R.~Landi,$^1$ N.~Masetti,$^1$ A.~De Rosa$^3$, 
\newauthor
  F.~Panessa,$^3$ M. Molina $^2$, A.~J.~Dean$^2$, M.~Perri$^4$, J.~Tueller$^5$\\
$^1$ IASF/INAF, via Gobetti 101, I-40129 Bologna, Italy,\\
$^2$ School of Physics and Astronomy, University of Southampton, SO17 1BJ, Southampton, U.K.\\
$^3$ IASF/INAF, via del Fosso del Cavaliere 100, I-00133 Roma, Italy \\
$^4$ ASI Science Data Center, ASDC c/o ESRIN, via G. Galilei snc I-00044 Frascati, Italy \\
$^5$ NASA/GSFC,code 661,Greenbelt, MD 20771 USA }

\begin{document}

\date{}

\pagerange{\pageref{firstpage}--\pageref{lastpage}} \pubyear{2008}

\maketitle

\label{firstpage}

\begin{abstract}
  Narrow Line Seyfert 1 galaxies (NLS1) are very interesting objects
  which display peculiar properties when compared to their broad line analogues (BLS1).
  Although well studied in many wavebands, their behaviour at $>$10 keV  is poorly studied
  and yet important to discriminate between models invoked to explain the 
  complexity observed in the X-ray band.  
  Here we present for the first time high energy observations (17-100 keV) of five NLS1 galaxies 
  (3 bona fide and 2 candidates) detected by  \emph{INTEGRAL}/IBIS and provide for all of them a broad band 
  spectral analysis using data obtained by \emph{Swift}/XRT below 10 keV.
  The combined \emph{INTEGRAL} spectrum is found to be steeper ($\Gamma$=2.6$\pm$0.3) than those
  of classical Seyfert 1 objects. This is due to a
  high  energy cutoff, which is required in some individual fits as in the average broad band spectrum.
  The location of this high energy cutoff  is at lower energies (E$\leq$60 keV) than typically seen
  in classical type 1 AGNs; a reflection component may also be present but its value (R$<$0.8) is compatible 
  with those seen in standard Seyfert 1s. 
  We do not detect a soft excess in individual objects but only in their cumulative spectrum.
  Our results suggest a lower plasma temperature for the accreting plasma  which combined to the high accretion rates
  (close to the Eddington rate) point to different nuclear conditions in broad and narrow line Seyfert 1 galaxies,
  likely related to different evolutionary stages. 
\end{abstract}


\begin{table*}
\begin{center}
\centerline{Table 1: INTEGRAL/ISGRI and Swift-XRT data}
\label{table:1}
 \begin{tabular}{l c c l l c c c}
  \hline\hline
Name               &     $z$ & FWHM(H$_\beta^\dagger$) & OIII/H$_\beta$ &FeII/H$_\beta$ &
M$_{BH}$ ($\times$ 10$^{7}$ M$_{\bigodot}$) &   N$_{H_{Gal}}^\ddagger$  & Ref. \\
\hline
IGR J14552-5133    & 0.0160 & 1700 & 0.70$\pm$0.07 & $\le$ 1.70     & 0.2 & 3.37 & [1,2] \\
IGR J16185-5928    & 0.0350 & 4000 & 0.20$\pm$0.03 & 0.70$\pm$0.06  & 2.8 & 2.48 & [1,2] \\
IGR J16385-2057    & 0.0269 & 1700 & 0.50$\pm$0.06 & 1.20$\pm$0.20  & 0.7 & 1.20 & [2,3,4,5] \\
IGR J19378-0617    & 0.0106 & 2700 & 0.52$\pm$0.04 & 0.89$\pm$0.06  & 0.3 & 1.47 & [6]\\
SWIFT J2127.4+5654 & 0.0147 & 2000 & 0.72$\pm$0.05 & 1.30$\pm$0.20  & 1.5 & 7.87 & [2,7] \\
\hline
 \hline
\end{tabular}
\end{center}
\small
$\dagger$:error on FWHM is typically 300 km/s;  $\ddagger$: $\times$ 10$^{21}$ cm$^{-2}$\\
Ref: [1] Masetti et al. 2006a; [2] this work; [3] Jones et al. 2004; [4] Masetti et
al., 2006b,
[5] Masetti et al. 2008; [6] Rodriguez-Ardila et al. 2000; [7] Halpern 2006.
\end{table*}

\section{Introduction}
Narrow Line Seyfert 1s (hereafter NLS1) 
are a sub-class of Seyfert 1 galaxies that display very peculiar and
interesting properties. At  optical wavelength they differ from Broad Line
Seyfert 1 (BLS1) galaxies for having: (1) the full width at half-maximum (FWHM) 
of the H$_\beta$ line lower than 2000 km s$^{-1}$ (but see section 2); (2) the
permitted lines only slightly broader than the forbidden lines; (3)
the [OIII] $\lambda$5007/H$_\beta$ ratio $<$3; and (4) unusually
strong FeII and other high ionisation emission line complexes
(Osterbrock \& Pogge 1985).  A popular explanation of the differences
in the optical properties across the Seyfert population is that NLS1s
have smaller black hole masses than their broad line analogues. 
Given that NLS1s have comparable luminosity to
that of the BLS1s (Pounds, Done \& Osborne 1995), they must be emitting at higher
fractions of their
Eddington luminosity; hence, higher fractional accretion rates ($\dot{m}$ =
$\dot{M}$/$\dot{M}_{Edd}$)
are also required.  
Alternative explanations for the difference observed across the Seyfert 
population include the possibility that NLS1s have more distant  broad-line regions (BLR) and, hence, 
smaller Keplerian line widths (Wandel 1997) 
or that NLS1s are observed preferentially close to face on. 
The former possibility is discounted since NLS1s and BLS1 have BLRs 
of comparable sizes
(Kaspi et al. 2000; Peterson et al. 2000); the latter is disfavoured by the analysis
of Boroson \& Green 
(1992) and the fact that the inner regions of BLS1s also appear to be observed close
to face-on (Nandra et al. 1997).  
The first and most plausible scenario suggests that black holes in NLS1 have not yet
been fed 
enough to become massive and are in a rapidly growing phase (Mathur 2000); 
if NLS1 are indeed in an early phase of black hole evolution, then they are key
objects for
studying formation and evolution of Active Galactic Nuclei (AGNs). 
Their detailed studies over many wavebands in search of similarities/differences
with their broad
line counterparts can enable us to reveal the formation mechanisms and processes of
central black holes 
in the local Universe and help understanding  QSO formation and evolution. 

The waveband where NLS1 show the most marked differences compared to broad line AGNs
is the X-ray band. In this waveband the best known properties of NLS1 are the presence of a
soft excess in most but not all objects (Turner, George \& Netzer, 1999), an unusually 
strong X-ray variability (Boller et
al. 1997; Leigly 1999) and 2-10
keV continuum slopes  steeper ($\Gamma$=2.19) than in normal Seyfert 1s ($\Gamma$=1.73, Veron-Cetty, Veron and Goncalves 2001).  
With the higher sensitivity of current X-ray telescopes came
the discovery of a sharp spectral drop at around 7 keV in a number of
NLS1 (see Gallo 2006 and references therein).  
The origin of this behaviour is still debated, but two models have been put forward to
explain this spectral complexity: partial covering (Tanaka et al.
2004) and reflection/light bending (Fabian et al 2002, Miniutti \& Fabian 2004). 
In the first case the drop is produced by absorption of the continuum by dense
material which
partly obscures the source; in the second case it is simply the blue wing of the
relativistically broadened iron line.  
Despite being successful in explaining most of the 2-10 keV spectral complexity
observed in NLS1, the first model does not describe the nature of the
primary continuum source nor provides physically acceptable parameters
for the soft excess component usually fitted with a black body model.
Reflection of the power law continuum source off the cold accretion
disc describes more adequately all the features seen in the X-ray
spectra of NLS1. Clearly broad band data extending above 10 keV are
crucial to discriminate between these two competing models;
unfortunately these types of observations are rare for NLS1, with only
a few broad band spectra being measured by \emph{BeppoSAX} (Comastri
et al 2001, Dadina 2007). In these few cases, data above 10 keV did
not provide conclusive results. Now, the number of known NLS1 bright
above 10 keV is increasing thanks to the \emph{INTEGRAL} and \emph{Swift}
all sky surveys (Bird et al. 2007, Tueller et al. 2007) thus providing the base for a
first study of their high energy (and consequently also broad-band)
spectral behaviour.  Here we present \emph{INTEGRAL} results for a set
of 5 newly discovered NLS1 (3 bona-fide and 2 candidates);
our data  suggest a steeper high energy
spectral shape with respect to broad line Seyfert galaxies and the
preference for  a lower cut-off energy ($\leq$60 keV). 
We do not detect soft excess emission
in individual objects but only in their cumulative spectrum, which also provides an upper limit on the average reflection 
parameter (R$<$0.8). All this observational evidence makes these objects 
interesting in their own right even if not all of them will be confirmed as NLS1.

\section{The Sample} 
In this work we present a sample of 5 NLS1  (3 bona-fide and
2 candidates): IGR J14552-5133, IGR J16185-5928, IGR J16385-2057,
IGR J19378-0617 and SWIFT J2127.4+5654.  All 5 objects are gamma-ray
selected AGNs discovered by \emph{INTEGRAL}/IBIS and Swift/BAT; all 5
are reported in the last \emph{INTEGRAL}/IBIS survey catalogue (Bird
et al. 2007). The large (arcmin) gamma-ray error box has been reduced
to arcsec uncertainty using X-ray follow-up observations performed by
the Swift/XRT telescope (Malizia et al. 2007, Rodriguez et al. 2008 and this work)
\footnote {For IGR J16185-5928 the X-ray position provided by
the XRT observations presented in this work (RA: 16$^{h}$ 18$^{m}$ 26.38$^{s}$, 
Dec: -59$^{\circ}$ 28$^{m}$ 45.27$^{s}$, 5 arcsec  uncertainty) is fully consistent with the ROSAT position.} 
The reduced X-ray error boxes
allowed the identification in each case of the optical counterpart
and, through spectroscopic observations, the classification of each
source. Details on the optical observations and results can be found
in Masetti et al. (2006a, 2006b, 2008), Rodriguez-Ardila et al. (2000)
and Halpern (2006).  To have more detailed information on SWIFT
J2127.4+5654, we have spectroscopically observed this source on 2007
May 13 with the `G.D. Cassini' 1.5m telescope of the Astronomical
Observatory of Bologna located in Loiano (Italy) plus BFOSC, using
Grism \#4 and a slit of width 2$\farcs$5, securing a 3500-8700
\AA~nominal spectral coverage. The reduction of these data was the
same as, e.g., that in Masetti et al. (2004); the resulting spectrum
confirms the previous classification as a NLS1 and  further provides
the information reported in Table 1.

Following Osterbrock \& Pogge (1985), NLS1s were initially defined as having the [OIII]
$\lambda$5007/H$_{\beta}$ ratio $<$3 and narrow "broad" Balmer
components (FWHM(H$_\beta$) $\le$ 2000 km s$^{-1}$). However, there is a
continuous distribution of optical line widths in Seyfert 1s and the
separation between BLS1s and NLS1s is somehow arbitrary. Sulentic et
al. (2000 and references therein) suggest that AGN properties change
more significantly at a broad line width of 4000 km s$^{-1}$ while
many objects in the literature are found to display NLS1 properties
(i.e. strong Fe II emission, a soft X-ray excess and variability)
despite the fact that their FWHM(H$_\beta$) exceeds 2000 km s$^{-1}$ (see
examples in Veron-Cetty, Veron and Goncalves 2001); on the basis of
these evidences, the latter authors suggested that a more meaningful
parameter to distinguish NLS1 from BLS1 is the ratio FeII/H$_{\beta}$,
where a cut at a value of 0.5 discriminates between objects with strong
and weak Fe II emission. Table 1 provides for all our sources the
relevant parameters for their classification as NLS1 (FWHM(H$_{\beta}$),
OIII/H$_{\beta}$ and FeII/H$_{\beta}$); it is evident from the listed
values that most of our objects fulfill at least 2 sometimes 3 of the
NLS1 definition requirements: all but two sources (IGR J16185-5928 and
IGR J19378-0617) have FWHM(H$_{\beta}$) below 2000 km s$^{-1}$, but all
comply with the Sulentic et al. (2008) less stringent limit on the broad
line width; all have an OIII/H$_{\beta}$ ratio below 3; only in 
IGR J14552-5133 the FeII bump around 4500 $\AA$ (restframe) is not
detected, but the limit on the FeII/H$_{\beta}$ ratio is not stringent
enough to exclude this object from the sample.  
Finally IGR J19378-0617 (also known as 1H1934-063) has already been classified in
the literature as a NLS1 (Rodriguez-Ardila et al. 2000, Nagao,
Marayama and Taniguchi 2001).  Some radio-loud AGN exhibit optical
spectra which display narrow permitted lines
and weak [OIII] emission and are so easily misclassified as NLS1.
Only IGR J16385-2057, IGR J19378-0617 and
Swift J2127.4+5654 are reported as radio sources with a 20 cm flux of
6.8, 42.2 and 6.4 mJy (Condon et al. 1998) but none is sufficiently
bright to be considered radio loud (see for example radio loudness
definition in Sulentic et al. 2008).  Overall we can conclude saying
that 3 objects (IGR J16385-2057, IGR J19378 and Swift J21247.4+5654) in
our sample are bona-fide NLS1 while the remaining 2 can be considered at
this stage only NLS1 candidates with IGR J14552-5133 probably a more
convincing case than IGR J16185-5928. Clearly more detailed optical
spectroscopy is necessary and encouraged to refine and confirm the
optical classification of these last two objects.  
To complete the set of information on our sources, we also report in Table 1 the Galactic
column density along the direction of each source (Dickey \& Lockman 1990) and the mass of the
central black hole.  For IGR J14552-5133, IGR J16185-5928 and IGR
J16385-2057 the masses have been calculated by Masetti et al. (2006a)
while for the remaining two objects they were estimated in the same way
following Wu et al. (2004) and Kaspi et al. (2000).  As expected the
masses of our NLS1 galaxies are small compared with the typical values
found in BLS1 (see for comparison black hole masses of BLS1
discovered by INTEGRAL in Masetti et al. 2006a and Masetti et
al. 2008).

\begin{table}
\begin{center}
\centerline{Table 2: INTEGRAL/ISGRI}
\renewcommand{\footnoterule}{}
\begin{tabular}{l r l c r c }
\hline\hline
Source             & Expo$^{1}$   & Count Rate$^{3}$ &  $\Gamma^{4}$  & $\chi^{2}$/dof &  F$^{4}$ \\
\hline
IGR J14552-5133    & 1966               & 0.12$\pm$0.02                 & 2.49$^{+0.74}_{-0.62}$ & 6.3/8           & 0.92 \\
IGR J16185-5928    & 1945               & 0.25$\pm$0.02                 & 2.36$^{+0.46}_{-0.41}$ & 8.3/8           & 1.70 \\
IGR J16385-2057    &  973               & 0.18$\pm$0.03                 & 3.12$^{+0.81}_{-0.69}$ & 4.0/8           & 1.16 \\
IGR J19378-0617    &  823               & 0.19$\pm$0.04                 & 2.59$^{+0.78}_{-0.66}$ & 16.3/8          & 1.45 \\
SWIFT J2127.4+5654 &  675               & 0.35$\pm$0.03                 & 2.73$^{+0.32}_{-0.29}$ & 12.0/8          & 2.46 \\
\hline
\end{tabular}
\end{center}
(1) Ksec; (2) cts/sec in 17-100 keV band; (3) in the 17-100 keV band; (4) flux in 20-100 keV band in units of 10$^{-11}$ erg cm$^{-2}$ 
\end{table}

\begin{table*}
\begin{center}
\centerline{Table 3: Swift/XRT}
\renewcommand{\footnoterule}{}
\begin{tabular}{l c r l c c r c }
\hline\hline
Source                 & date     & Expo$^{1}$ & Count Rate$^{2}$ &  $\Gamma^{3}$  & N$_{H}^{4}$ & $\chi^{2}$/dof &  F$^{5}$ \\
\hline
IGR J14552-5133 (obs1) & 27-12-06 & 3728             & 0.21$\pm$0.007         & 1.84$\pm$0.11          &   -     &    34/35        & 0.99 \\
IGR J14552-5133 (obs2) & 02-01-07 & 4695             & 0.20$\pm$0.006         & 2.08$\pm$0.10          &   -     &    32/43        & 0.76 \\
IGR J16185-5928 (obs1) & 09-05-08 & 4145             & 0.12$\pm$0.005         & 1.77$\pm$0.15          &   -     &    9/22         & 0.56 \\
IGR J16185-5928 (obs2) & 12-05-08 & 5556             & 0.09$\pm$0.004         & 2.09$\pm$0.16          &   -     &    24/21        & 0.30 \\
IGR J16185-5928 (obs3) & 16-05-08 & 5171             & 0.17$\pm$0.005         & 2.04$\pm$0.16          &   -     &     36/39       & 0.59 \\
IGR J16385-2057 (obs1) & 07-10-07 & 4540             & 0.22$\pm$0.007         & 2.11$\pm$0.18          & 0.12$\pm$0.06 & 43/43     & 0.71 \\
IGR J16385-2057 (obs2) & 08-10-07 & 4592             & 0.24$\pm$0.007         & 2.18$\pm$0.17          & 0.11$\pm$0.05 & 43/46     & 0.70 \\
IGR J19378-0617 (obs1) & 26-09-07 & 2492             & 0.51$\pm$0.014         & 2.48$\pm$0.08          &   -      &    46/55       & 4.90 \\
IGR J19378-0617 (obs2) & 26-11-07 & 7214             & 0.44$\pm$0.007         & 2.46$\pm$0.05          &   -      &    114/117     & 2.00 \\
SWIFT J2127.4+5654 (obs1) &  20-10-05 & 1726         & 0.27$\pm$0.012         & 1.96$\pm$0.18          &   -      &    35/41       & 1.53 \\
SWIFT J2127.4+5654 (obs2) &  03-12-05 & 8403         & 0.38$\pm$0.007         & 1.89$\pm$0.06          &   -      &    141/129     & 2.25 \\
\hline
\end{tabular}
\end{center}
(1) sec; (2) cts/sec in 0.1-6 keV band; (3) in the 0.1-6 keV band; (4) Intrinsic absorption in units of 10$^{22}$ cm${-2}$; \\
(5) flux in 2-10 keV band in units of 10$^{-11}$ erg cm$^{-2}$ 
\end{table*}

\section{INTEGRAL/IBIS data}
The \emph{INTEGRAL} data reported here come from of several pointings
performed by the IBIS instrument from revolution 12 (i.e., the first useful revolution) to revolution 520,
and the corresponding total exposures are given in Table 2.   
The maximum source significance in a single IBIS pointing or science window (sw) is quite low and consequently it is not possible to extract a good spectrum from any 
individual measurement. Therefore, for each source, all available sw's were processed with the Off-line Science Analysis (OSA) software (Goldwurm et al. 2003) v.5.1,
to produce images in 10 energy channels spanning the 17-100 keV range (channels over this band were logarithmically spaced to evenly distribute the counts) and 
the weighted average flux of the source in each energy channel was calculated from the individual OSA flux and variance images. This was then used to construct 
a standard spectral \emph{pha} file. A rebinned \emph{rmf} file was produced from the standard IBIS spectral response file to match the 10 
chosen energy channels. This method for producing the spectra of weak persistent sources with  \emph{INTEGRAL} data by imaging in fine energy bands is 
an established method and has been extensively tested in comparison with the standard analysis for bright sources.

Here, and in the following, the spectral analysis was performed with XSPEC v.11.3.2 and errors are
quoted at 90\% confidence level for one parameter of interest
($\Delta\chi^{2}=2.71$). 
Due to the low signal to noise ratio of these detections, a simple power law has been employed to fit the data.  
The results of the \emph{INTEGRAL} data analysis including count rates, power law photon indices, $\chi^{2}$ and flux in
the 20-100 keV band are all listed in Table 2. Despite the large uncertainties, the $\Gamma$ values measured by  \emph{INTEGRAL} 
are found to be systematically steeper than those typically observed in BLS1 galaxies. 
To verify this result and in view of the low statistical quality of the IBIS detections,
all 5 NLS1 have been compared with a set of 34 BLS1 belonging to a complete sample of  \emph{INTEGRAL/IBIS} AGNs selected in the 20--40 keV band 
(Malizia et al. in preparation). 
Figure 1 shows the histogram of the IBIS photon indices relative to BLS1 (open bins) and NLS1 (filled bins):
it is evident that the NLS1 have steeper spectra. A confirmation of this result comes from the KS-test which
gives a probability of only 0.006 that the two samples belong to the same population.

\begin{small}
\begin{figure}
\centering
\includegraphics[width=1.05\linewidth]{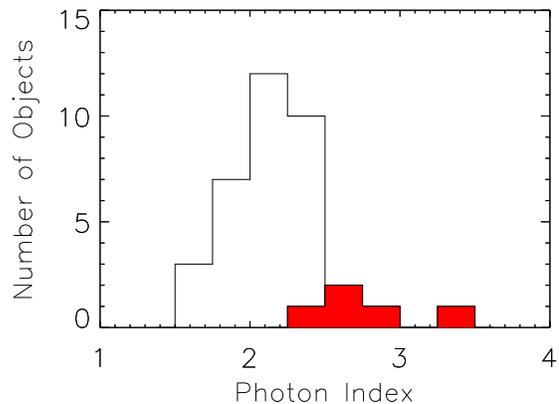}
\caption{Histogram of the INTEGRAL 17-100 keV photon indices of BLS1 galaxies together with
those of NLS1 galaxies (filled bins). }
\label{fig1}
\end{figure}
\end{small}

To further study this difference, we have fitted the 5 NLS1 spectra together and compared
them to the combined INTEGRAL spectral data of 5 BLS1 having
similar 20-100 keV fluxes and  exposures (ESO 209-12, FRL 1146, IGR J16558-5203,
IGR J17418-1212, SWIFT J1038.8-4942, Panessa et al. 2008, Landi et al. 2007a).  
The result of this comparison is shown in figure 2, where the contour plots of
normalisation versus photon index are shown for the sample of NLS1 and the set of BLS1: a
clear dichotomy is evident, with the NLS1 having a power law photon
index ($\Gamma$=2.60$\pm$0.30) steeper than the canonical
value ($\Gamma$=2.02$\pm$0.16) observed in  BLS1.
A steeper spectrum is also compatible with the limited number of NLS1 so far detected by \emph{INTEGRAL}. 
The fraction of NLS1 (vs BLS1) galaxies in optically selected samples is typically 15$\%$ but may increase to 20$\%$ 
dependingon their luminosity;  soft X-ray selected samples provide an even higher fraction  with, for example, 46$\%$ 
of NLS1 found among broad line AGN detected by ROSAT  (see Komossa 2008 and references therein).
Within the total sample of around 60 broad line AGN so far detected by
\emph{INTEGRAL} (Bird et al. 2007) only 5 (or $\sim$8$\%$ \footnote{Note that the same percentage is found in the complete
sample of  \emph{INTEGRAL/IBIS} AGNs.}) are possibly NLS1 against a greater number 
(from around 10 up  to 28) expected on the basis of the fractions seen in  optical or X-ray samples.

To further understand the spectral shape difference between NLS1 and BLS1, we have also fitted the NLS1 average spectrum with a 
$\Gamma$ = 2.02 power law (i.e. the same found for BLS1) and inspected the residuals with respect to this fit ($\chi^{2}$ =138.5 for 49 d.o.f):
Fig. 3 indicates a possible excess emission around 20-30 keV, resembling a reflection bump, and a deficit of counts at high energies, likely
due to an exponential cut-off in the power law continuum. While the  pexrav model would be more appropriate to describe these features, 
it is too sophisticated for the current low statistical significance of the  \emph{INTEGRAL} data. We have therefore employed a more elementary 
approach and tested the data against simpler  models such as a cut-off power law  and a broken power law: the first in search of a high 
energy exponential drop, the second to test for the presence of a reflection component since its peak should provide a flatter power 
law than at high energies and a break at around 30-40 keV. Keeping $\Gamma$ =2.02, the first model gives a cut-off at E$_c$=52$^{+29}_{-15}$ 
keV ($\chi^{2}$ =111.2 for 48 d.o.f) while the  second indicates  a break  at E=28$^{+7}_{-18}$ keV and a power law steepening to 
$\Gamma$  =3.03$^{+0.78}_{-0.43}$ ($\chi^{2}$ =110.2 for 47 d.o.f). In either case, there is a significant improvement (99.99$\%$)
with respect to the canonical AGN power law and this is therefore an indication of the reality of these extra high energy  features.
Although the \emph{INTEGRAL} data alone provide general information on the NLS1
high energy shape, they are not sufficient to constrain the spectral parameters of each source; clearly broadband data are needed for this purpose.

\begin{small}
\begin{figure}
\centering
\includegraphics[angle=-90,width=0.8\linewidth]{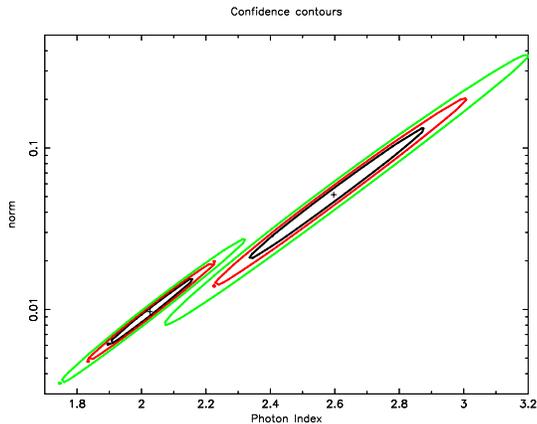}
\caption{Confidence contours (68\%, 90\%, 99\% levels) of the power law parameters of the set of BLS1 (left) and the NLS1 (right).}
\label{fig1}
\end{figure}
\end{small}

\begin{small}
\begin{figure}
\centering
\includegraphics[angle=-90,width=0.8\linewidth]{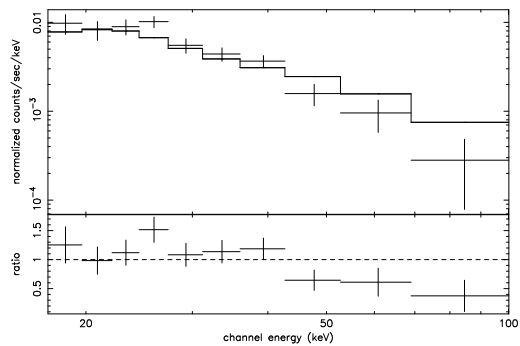}
\caption{17-100 keV average spectrum of INTEGRAL NLS1  fitted with a simple
power law with photon index fixed to 2.02 as in BLS1.}
\label{fig2}
\end{figure}
\end{small}

\section{Broad band data analysis}
\subsection{Individual sources}
All five NLS1 in the sample, 
have also been observed  more than once in the X-ray band with the
XRT telescope on board \emph{SWIFT}. 
The XRT data analysis of most of our sources has already been presented in two recent
papers by Malizia et al. (2007) and Rodriguez et al. (2007);  here for the first time we present the X-ray spectrum
of IGR J16185-5928. For this work the whole data set has been analysed in a uniform way,
following the procedures described in Malizia et al. (2007). Spectra have been fitted in the 0.1-6 keV band
using a simple power law absorbed by the Galactic column density, unless otherwise required.
Information related to these XRT measurements, such as observation date, net on source exposure, count rate, power law photon index, $\chi^{2}$/dof
values  and 2-10 keV fluxes are all reported in Table 3. 
All spectra are well fitted with  power laws having indices in 
the range 1.8-2.5 (mean 2.1) compatible with the values reported for NLS1 over the 2-10 keV band (Veron-Cetty, Veron and Goncalves 2001).
IGR J16385-2057  is the only source which requires absorption in excess to the Galactic value at a level of 10$^{21}$ cm$^{-2}$. 
X-ray flux variability, a typical feature of  NLS1 is observed in four out of five objects with changes greater than a factor of 1.3 and up to 2.4;
variations are seen on short (days) and long (months) timescales. \\
In none of our objects we find evidence for strong soft X-ray emission but this maybe due to the low statistical quality of the X-ray data.
Furthermore, a strong soft excess component is not a universal characteristic of NLS1 since many exhibit weak 
and/or low-$\Gamma$  X-ray emission (see a few examples described in  Veron Cetty-Veron, Veron and Goncalves 2001). 
Williams, Mathur and Pogge (2004) have in fact shown that the soft X-ray selection of NLS1 introduces a strong bias towards objects with 
an ultrasoft excess and demonstrate that the same 
is not true for optically selected samples. 
Also in the prototype NLS1, IZw1, a weak soft excess has been recently found and discussed by Gallo et al. (2007).
Our objects are hard X-ray selected and may belong to the
subset of NLS1 characterised by weaker soft X-ray emission.

Comparing Table 2 and 3,  it  is  clear that there is a change of photon indices going from the  X-ray data to the
\emph{INTEGRAL}/IBIS one, suggesting a more complex shape than a simple power law. 
Since  \emph{INTEGRAL} represents an average source state and in view of the low statistical quality of the XRT data,
we have combined all available X-ray observations of each NLS1 to produce an average 2-10 keV spectrum; this has then been fitted in
combination with  \emph{INTEGRAL/IBIS} data. 
Following the elementary approach adopted in section 3, 
the broad band continuum of each object has been fitted first 
with a cut-off power law to search for a high energy break and then  with a broken power law
to consider the possible presence of a reflection component. 
A constant has been introduced in the fits to account for flux/calibration  mismatch between XRT and IBIS.
Note however that this constant has been found to be close to one in non variable sources (see for example Landi et al. 2007b),
indicating that higher or lower values are most likely related to flux variations.
From this analysis we found that a high energy cut-off and a reflection hump is required by
the data in IGR J16385-2057 and SWIFT J2127.4+5654 at 99.99\% and 98\%/97\% respectively;
the cutoff lies in the range  23-52 keV while the broken power law break occurs at 
at around 30 keV (see Table 4) as expected in the presence of the a reflection component.  

\subsection{Average Spectrum}
In order to improve the statistics and provide better constrains on both cut-off and reflection parameters, 
we have then used the broad band spectra of our 5 sources summed together.
When we fit this average 0.1-100 keV spectrum with an absorbed power law model, we
get a value of N$_{H}$ $\sim$ 2 $\times$ 10$^{21}$ cm$^{-2}$ compatible with the mean of individual column densities (Galactic and intrinsic), 
a photon index $\Gamma$=1.82$^{+0.04}_{-0.04}$
and a good value for the cross calibration constant between XRT and ISGRI (C=1.09$^{+0.17}_{-0.14}$)  
but the fit is rather poor ($\chi^{2}_{\nu}$=458/336).
This is due to deviations from the power law at low ($<$0.5 keV) and high energies ($>$20 keV).
We cured the first introducing a black body component which provides a fit improvement greater than 99.99\% (see Table 5) and 
a kT of around 70 eV slightly lower but still compatible with the average seen in NLS1 (0.1-0.15 keV, Crummy et al. 2006).
This result indicates that a soft excess is present in our sources but is less prominent than 
in more classical (and soft X-ray selected) NLS1; it is possible that the hard X-ray observations
as the optical ones select sources with weaker soft X-ray emission.
At high energies, the addition of a high energy cut-off is equally 
required with a high significance  ($>$99.99\%, $\Delta \chi$ = 41 for 1 degree of freedom with respect to the black body plus
power law fit) and gives a cut-off energy at E$_c$=38$^{+17}_{-10}$ keV.
The broad band spectrum and the confidence contours of high energy cut-off
versus photon index using this model are shown in figures 4 and 5 respectively.

Although the broken power law provides an improvement with respect to the black body plus power law model
and a well constrained break in the 20-40 keV range,
it gives a worse fit than the model with the high energy cutoff suggesting a less prominent contribution of the reflection
component. To further check this results we employ the  \texttt{pexrav} model to our broad band average spectrum;
this gives similar parameters values of the  \texttt{bb+cutoffpl} model (see Table 5), no improvement in the fit but
an upper limit on the reflection of R$<$0.8.
Therefore, the conclusion of  our broad band  analysis is that a high energy  cut-off is strongly required by the data 
and that it is located  at energies  $\lesssim$ 60 keV (see figure 5);
a reflection component may also be present but its value is compatible with values observed in the BLS1.

Closely linked to our analysis is also the estimate for  each source of the accretion rate relative to the Eddington rate, i.e. the ratio of  
L$_{bol}$/L$_{Edd}$ where L$_{bol}$ is estimated here from the 1-100 keV luminosity plus a fixed bolometric correction 
(taken here as L$_{1-100 keV}$ = 4.2\%L$_{bol}$, Risaliti \& Elvis 2004)
and L$_{Edd}$=1.3 $\times$10$^{38}$M/M$_{\bigodot}$. We obtained 1.3, 0.5, 1.2, 0.9, 0.4 for IGR J14552-5133, IGR J16185-5928, IGR J16385-2057, IGR J19378-0617
and SWIFT J2127.4+5654 respectively; these values support the claim that our NLS1 accrete close to  the Eddington rate,
as expected (e.g. Boroson \& Green 1992).

\begin{table*}
\begin{center}
\centerline{Table 4: IBIS/ISGRI + Swift/XRT broad band data analysis}
\renewcommand{\footnoterule}{}
\begin{tabular}{r c c c c c c}
\hline\hline
\multicolumn{7}{c}{\bf {IGR J14552-5133 - sum}}\\
\hline
model              &      $\Gamma_{1}$       &  E$_{b}$          & $\Gamma_{2}$        & E$_{c}$  &  Const.                & $\chi^{2}$/dof \\
\hline
power law          & 1.93$^{+0.09}_{-0.08}$  &     -             &     -               &  -                & 0.97$^{+0.37}_{-0.29}$ & 79/86 \\
cutoffpl           & 1.88$^{+0.08}_{-0.09}$  &     -             &     -               & 46$^{+255}_{-20}$ & 1.75$^{+1.47}_{-0.78}$ & 75/85 \\
bknpo              & 1.92$^{+0.07}_{-0.07}$  & 50$^{+18}_{-22}$  &    \it{unconstr.}  &         -         & 1.13$^{+0.41}_{-0.33}$ & 74/84 \\
\hline
\multicolumn{7}{c}{\bf {IGR J16185-5928 - sum}}\\
\hline
model              &      $\Gamma_{1}$       &  E$_{b}$          & $\Gamma_{2}$        & E$_{c}$           &  Const.                & $\chi^{2}$/dof \\
\hline
wa power law       & 1.96$^{+0.07}_{-0.07}$ &     -              &     -               &     -             & 3.45$^{+1.00}_{-0.78}$  & 78/91  \\
wa cutoffpl        & 1.94$^{+0.07}_{-0.07}$ &     -              &     -               & $>$46             & 4.21$^{+2.00}_{-1.10}$  & 77/90  \\
wa bknpo           & 1.95$^{+0.07}_{-0.07}$ & \it{unconstr.}     & \it{unconstr.}      &         -         & $>$3                    & 74/89 \\
\hline
\multicolumn{7}{c}{\bf {IGR J16385-2057$^{\dagger}$ - sum}}\\
\hline
model              &      $\Gamma_{1}$       &  E$_{b}$          & $\Gamma_{2}$        & E$_{c}$           &  Const.                & $\chi^{2}$/dof \\
\hline
wa power law       & 2.15$^{+0.12}_{-0.12}$ &     -              &     -               &     -             & 2.71$^{+1.43}_{-0.95}$  & 97/96  \\
wa cutoffpl        & 2.03$^{+0.15}_{-0.15}$ &     -              &     -               & 29$^{+85}_{-15}$  & 5.86$^{+5.70}_{-2.90}$    & 91/95  \\
wa bknpo           & 2.12$^{+0.12}_{-0.12}$ & 27$^{+13}_{-22}$   & $>$2.6              &         -         & 3.79$^{+31}_{-1.36}$    & 90/94 \\
\hline
\multicolumn{7}{c}{\bf {IGR J19378-0617 - sum  }}\\
\hline
model              &      $\Gamma_{1}$       &  E$_{b}$          & $\Gamma_{2}$        & E$_{c}$           &  Const.                & $\chi^{2}$/dof \\
\hline
power law          & 2.44$^{+0.04}_{-0.04}$ &    -              &       -              & -                 &  1.65$^{+0.55}_{-0.49}$ & 163/177 \\
cutoffpl           & 2.43$^{+0.04}_{-0.05}$ &    -              &                      &     $>$35         &  1.94$^{+1.52}_{-0.61}$ & 162/176 \\
bknpo              & 2.44$^{+0.04}_{-0.04}$ & 41$^{+41}_{-28}$   & \it{unconstr.}      &    -              &  1.93$^{+0.63}_{-0.56}$ & 160/175 \\ 
\hline
\multicolumn{7}{c}{\bf {SWIFT J2127.4+5654 - sum }}\\
\hline
model              &      $\Gamma_{1}$       &  E$_{b}$          & $\Gamma_{2}$        & E$_{c}$           &  Const.                & $\chi^{2}$/dof \\
\hline
power law          & 1.92$^{+0.05}_{-0.05}$ &    -              &   -                  &    -              &  1.09$^{+0.23}_{-0.19}$ &  182/161  \\
cutoffpl           & 1.81$^{+0.06}_{-0.06}$ &    -              &   -                  & 33$^{+19}_{-10}$  &  2.15$^{+0.69}_{-0.52}$ &  153/160 \\
bknpo              & 1.89$^{+0.05}_{-0.05}$ & 30$^{+6.3}_{-6.3}$ & 3.78$^{+2.14}_{-0.88}$ &    -           &  1.38$^{+0.28}_{-0.23}$ &  151/159 \\ 
\hline
\end{tabular}
\end{center}
$\dagger$: this is the only source in which the data required an extra absorption in addition to the Galactic
column density of about 1.2 $\pm$ 0.05 $\times$ 10$^{22}$ cm$^{-2}$
\end{table*}

\begin{small}
\begin{figure}
\centering
\includegraphics[angle=-90,width=0.8\linewidth]{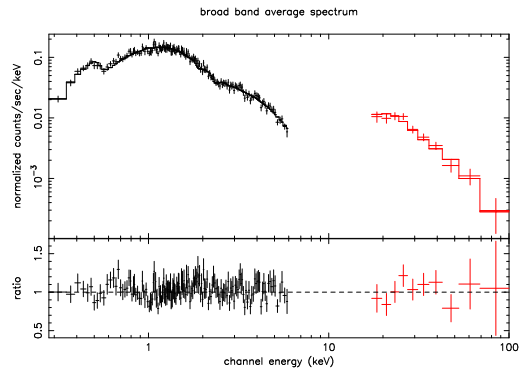}
\caption{Broad band (0.1-100 keV) average data fitted with a blackbody + cut-off power law model.}
\label{fig2}
\end{figure}
\end{small}

\begin{small}
\begin{figure}
\centering
\includegraphics[angle=-90,width=0.8\linewidth]{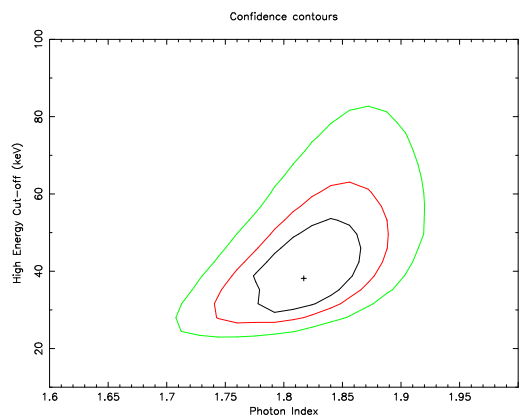}
\caption{Contours of the cut-off energy versus photon index at 68\%, 90\% and 99\% confidence levels, using broad band
XRT-IBIS average data.}
\label{fig2}
\end{figure}
\end{small}

\section{Discussion and Conclusions}
The 5 AGN discussed in this work have optical characteristics
which indicate that 3 are bona fide NLS1 while 2 are only candidates  
and as such provide the first sample (albeit
small) of such objects selected above 10 keV. Analysis of their high
energy spectra reveals for the first time a different behaviour above
10 keV with respect to their broad line analogues and in particular a
steeper ($\Gamma$=2.6) power law slope than the canonical $\Gamma$
value of $\sim$ 2 associated to standard Seyferts.  Based on our analysis, the steeper primary continuum is likely
due to a lower  energy cut-off ( E$_{c}$ $\le$ 60 keV) than typically observed in BLS1 (Malizia et al. 2003)
while the presence of a  Compton reflection component is not 
strongly required by the data and in any case the value found (R$<$0.8)
is compatible with those generally observed in broad line AGNs. 
This observational evidence makes the present sample apart from the standard type 1 AGN
discovered by IBIS even if not all will be confirmed NLS1.

Modelling of classical Seyfert spectra ascribe the power
law to the inverse Compton scattering of soft photons off hot
electrons. Variations to this baseline model depend on the energy
distribution of these electrons and their location in relation to the
accretion disc, often a hot corona above the disk. 
Within this model the power law
photon index and cut-off energy are directly related to the
temperature and optical depth of the Comptonising hot plasma so that
precise information on these two parameters provide clues for our
understanding of the primary emission mechanism, geometry of the
source and physics of the plasma near the central powerhouse. If as
suggested,  NLS1 galaxies are evolutionarily young objects, powered by
the accretion of gas onto central black holes that are lower in mass and accreting at a higher rate
than their broad-line counterparts, measurements of a different high energy continuum
may have strong implications on the cosmic evolution of
Supermassive Black Holes. Using appropriate relations (Petrucci et al. 2001) it is possible to 
derive from the values of $\Gamma$ and E$_c$, estimates of the temperature
kTe and optical depth $\tau$ of the Comptonising hot plasma;
in the case of NLS1 galaxies we obtain that kTe$\sim$ 10-20 keV and $\tau \gg$1, i.e.
a much lower temperature than typically found in BLS1 but similar plasma
thickness. It can therefore be argued that the
Comptonising hot plasma of NLS1 did not have enough time to reach the
condition of the older, more massive and lower rate accreting black holes found in BLS1.
This maybe related to the higher accretion rates associated to NLS1,
which  provide the conditions 
for a much higher  cooling of the emitting plasma and therefore a lower temperature. 
In fact, the accretion rates estimated for  our sources are  indeed close the Eddington rate.

Clearly, optical spectra with higher signal to noise ratio are desirable  (and encoraged) to firmly assess the NLS1 classification
in particular in the case of IGR J14552-5133 and IGR J16185-5928.
Furthermore, to better characterise our sources, we need higher quality X-ray data as 
available from XMM observations and possibly simultaneous broad band coverage
as it is possible  with \emph{Suzaku}.
We hope in the near future to be able to obtain both these type of measurements and put 
stronger constraint on the cut-off energy and reflection
components in individual objects. It is also important to properly classify
all AGN detected by \emph{INTEGRAL}/IBIS and \emph{SWIFT}/BAT in search for new
gamma-ray selected NLS1, which can be tested against  our findings.

\begin{table}
\begin{center}
\centerline{Table 5: IBIS/ISGRI + Swift/XRT  AVERAGE SPECTRUM}
\renewcommand{\footnoterule}{}
\begin{tabular}{l c c c c}
\hline\hline
                 & \texttt{mo po}         & \texttt{mo bb po}      & \texttt{mo bb cutoffpl} & \texttt{mo bb bknpo}    \\
\hline
N$_{H}^{\dagger}$& 0.19$^{+0.01}_{-0.01}$ & 0.27$^{+0.04}_{-0.06}$ & 0.21$^{+0.02}_{-0.02}$ & 0.23$^{+0.05}_{-0.01}$ \\
$\Gamma$         & 1.82$^{+0.04}_{-0.04}$ & 1.97$^{+0.05}_{-0.04}$ & 1.82$^{+0.06}_{-0.06}$ & 1.90$^{+0.05}_{-0.05}$ \\
kT               &    -                   & 0.07$^{+0.03}_{-0.01}$ & 0.05$^{+0.03}_{-0.02}$ & 0.06$^{+0.03}_{-0.01}$ \\
E$_{cutoff}$     &    -                   &     -                  & 38$^{+17}_{-10}$       &            -           \\
E$_{break}$      &    -                   &     -                  &     -                  & 27$^{+7}_{-13}$        \\
$\Gamma_{2}$     &    -                   &     -                  &     -                  & 3.02$^{+0.70}_{-0.43}$ \\
C                & 1.09$^{+0.17}_{-0.14}$ & 1.43$^{+0.22}_{-0.08}$ & 2.33$^{+0.53}_{-0.45}$ & 1.60$^{+0.27}_{-0.24}$ \\
$\chi^{2}$/dof   & 458/336                & 438/334                & 397/333                & 400/332                \\
\hline
\hline
\end{tabular}
$\dagger$ in units of 10 $\times$ 10$^{22}$ cm$^{-2}$\\
\texttt{po}: power law
\texttt{bb po}: blackbody + power law\\
\texttt{bb cutoffpl}: blackbody + cut-off power law\\
\texttt{bb bknpo}: blackbody + broken powe law     
\end{center}
\end{table}

\section*{Acknowledgements}
We acknowledge ASI  financial and programmatic support via contracts I/008/07/0 and I/088/06/0. 
Optical data used  are partially based on observations performed with Loiano  Astronomical
Observatory (Bologna, Italy). We thank Antonio De Blasi for the night assistance at the
telescope.

\end{document}